# DYNAMICS OF INHOMOGENEOUS POPULATIONS AND GLOBAL DEMOGRAPHY MODELS

Georgy P. Karev

Oak Ridge Institute for Science and Education (ORISE)
National Institute of Health
E-mail: gkarev@hotmail.com

**Abstract**.

The dynamic theory of inhomogeneous populations developed during the last decade predicts several essential new dynamic regimes applicable even to the well-known, simple population models. We show that, in an inhomogeneous population with a distributed reproduction coefficient, the entire initial distribution of the coefficient should be used to investigate real population dynamics. In the general case, neither the average rate of growth nor the variance or any finite number of moments of the initial distribution is sufficient to predict the overall population growth. We developed methods for solving the heterogeneous models and explored the dynamics of the total population size together with the reproduction coefficient distribution. We show that, typically, there exists a phase of "hyper-exponential" growth that precedes the well-known exponential phase of population growth in a free regime.

The developed formalism is applied to models of global demography and the problem of "population explosion" predicted by the known hyperbolic formula of world population growth. We prove here that the hyperbolic formula presents an exact solution to the Malthus model with an exponentially distributed reproduction coefficient and that "population explosion" is a corollary of certain implicit unrealistic assumptions. Alternative models of world population growth are derived; they show a notable phenomenon, a transition from protracted hyperbolical growth (the phase of "hyper-exponential" development) to the brief transitional phase of exponential growth and, subsequently, to stabilization. The model solutions are consistent with real data and produce relatively accurate forecasts.

## 1. The statement of a problem and the main results

The simplest Malthus model of population growth is rather unrealistic on a large time scale as it predicts that the population would grow exponentially. However, the actually observed growth estimates for the total world population look even more unbelievable. Specifically, the growth of the world population $N$ over hundreds of years, up to ~1990, is described with high accuracy by the hyperbolic law

$$N(t)=C/(T-t)^k, \qquad (1)$$

where $C \approx 2*10^{11}$, $T \approx 2025$, $k \approx 1$ [4].

Formula (1) with $k=1$ is a solution to the model problem with a square law of growth:

$$dN/dt = N^2/C. \qquad (2)$$

Although equations (1), (2) are very simple analytical models, they fit well the past growth of the world population and so might reflect its real trend. Then, the human population development looks dramatically different from that of other biological populations in a free regime; in particular, the human population was considered as the only one with a positive feedback between the average reproduction coefficient (growth rate per individual) and the population size ([13, ch.21]). The analysis of simple conceptual models (1), (2) is a promising way to realize some basic principles of population dynamics.

From the "physical" point of view, equations (1) and (2) describe the self-similar non-linear growth of a statistically uniform system [6]. From a "biological" point of view, the growth rate in equation (2) is proportional to the number of "pair contacts" in a population. When applied to the world population, this means that the growth rate is proportional to the number of "pair contacts" in the *whole* of humankind and the individual reproduction coefficient is proportional to the world population. This fact is difficult to interpret from the point of view of elementary processes and it seems evidently wrong even for small populations. In addition, Eq. (2) and its mathematical corollary (1) predict a *demographic explosion* at some point in time as we approach the year $T \cong 2025$. This means that $N(t) \to \infty$ when $t \to T$, and the same is true for the population growth rate and individual reproduction coefficient. Thus, formulas (1) and (2) have no plausible interpretations near $t=T$ and, accordingly, a contradiction exists between the good numerical accuracy of formula (1) and the interpretation of Eq. (2).

Let us note that, usually, Eqs. (1) and (2) are considered to be equivalent; this assertion was the basis for some interesting attempts to improve the model (1) (see, e.g., [14]) or even to develop a new phenomenological demographic theory [6]. In the latter theory, the underlying assumption of the humankind uniformity and the interpretation of equation (2) in terms of the "information community" hypothesis are, perhaps, unrealistic, especially when dealing with pre-historic human populations that were composed of non-interbreeding subgroups. Let us emphasize that, paradoxically (from the standpoint of the theory developed in [6]), the hyperbolic growth was observed in the past but has substantially slowed down during the last decades when most of the world population became an "information community".

An alternative approach to understand the origin and boundaries of validity of formula (1) is suggested in this work. Most growth models suppose that all individuals in a population have identical attributes, primarily, identical rates of growth, death, and birth. Well known basic population models of Malthus, Verhulst, and Allee (see, e.g,, [9]) belong to these types of models. This assumption simplifies computation at the cost of realism; in the prediction of population growth, recognition of subgroups that vary in the rate of increase, can produce a higher population size than the projection of the whole population at the average rate of increase [12]. Furthermore, recognition of population heterogeneity can lead to unexpected effects [13], [15].

Let us assume that every individual possesses its *own value* of a parameter *a*, which describes its invariable property (such as hereditary attributes or a specificity of the local habitat). The parameter remains unchanged for any given individual and varies from one individual to another, so the population is not uniform. Any changes of mean value, variance and other characteristics of the parameter distribution with time are caused only by variation of the population structure. In the models that follow, we suppose that every individual possesses its own hereditary value of the reproduction coefficient, which is distributed over the population.

We explore the empirical hyperbolic dependence (1) within the framework of non-homogeneous population models and show that formula (1) presents a solution of the *Malthus model* with an *exponentially distributed* (rather than constant) Malthusian parameter which assumes values in the range [0,∞) and has initial mean and variance equal to 1/*T*. Thus, the hyperbolic dependence (1) is *not equivalent* to Eq. (2). The Malthusian growth of an inhomogeneous population in a free regime is a more plausible explanation of the initial hyperbolic growth. Below we show that a "hyper-exponential" (in particular, hyperbolic) growth is a necessary initial phase of the development of inhomogeneous populations. Let us underline that the inhomogeneous Malthus model shows the hyperbolic solution (1) if and only if the reproduction coefficient is distributed over the population according to the exponential distribution. These facts do not "justify" formula (1) but clarify the implicit assumptions that result in unrealistic prognoses given by this formula and Eq.(2).

The Malthusian parameter *a* must be biologically realistic and hence bounded, $0 \leq a \leq c = const$ (in contrast with the model (1)). It is proven below that the solution of an inhomogeneous Malthus model with a *bounded* exponentially distributed parameter is *finite at any instant.* According to this model, the population develops hyperbolically for a very long time and then, after a short transition period, changes the growth regime to an exponential one. To fit the demographic data for several thousand years until the end of the 20$^{th}$ century, one should assign $c \approx 0.1$. Therefore, the "demographic explosion is a corollary of the implicit (and obviously wrong) assumption that the individual reproduction coefficient may assume *unboundedly large* values with *positive* (however small) probabilities.

The transition from a Malthusian type of model to a logistic one with a *distributed* growth parameter, results in the demographic model with a bounded solution having the hyperbolic-like initial growth (until about 1980-1990). It is possible to reach a satisfactory agreement of the model solution with some existing forecasts (UNO and IIASA) and simultaneously with the world population data during the last 2000 years. The solutions to three derived models as well as the dynamic properties of mean values and variances of the growth parameter are compared. Further investigation based on the logistic and Allee type of models with distributed parameters is also promising.

The mathematical foundation of the approach described here is the theory of inhomogeneous population models [1], [7]. In this work, we use the approach developed in [8], [2] and briefly presented in section 2 (which may be useful for those whose goal is to apply mathematical models in biology and could be skipped by biologically-inclined readers). Section 2 contains necessary references for the main assertions in the core section 3 devoted to global demography models.

## 2. Mathematical foundations: inhomogeneous population models

### 2.1. Individual-based population model with a distributed parameter

The dynamic model of population $N(t)$ characterized by the growth factor $F(N,a)$ that depends on the parameter $a$ is given by the equation

$$dN/dt = NF(N,a). \qquad (3)$$

From now on we assume that every individual possesses its *own value* of the parameter $a$. Let us call an $a$-group a set of all individuals having a given value of the parameter $a$; then let $l(t,a)$ be the size of the $a$-group at some instant $t$. Using Eq. (3), we suppose that the growth factor of the $a$-group is $F(N,a)$. This means that the factor depends on the «group» parameter value and on the whole population size $N(t)$ but does not depend on the sizes of other groups. The following model (see [1], [7]) can describe the dynamic behavior of such a population:

$$dl(t,a)/dt = l(t,a) F(N,a), \qquad (4)$$

$$N(t) = \int_A l(t,a) da$$

where $A$ is a set of possible values of $a$, $l(0, a) = l_0(a)$ for all $a \in A$ and $N_0 = N(0) = \int_A l_0(a) da$ are given. Suppose that the growth factor $F(N,a)$ takes the form

$$F(N,a) = f(N) + a g(N). \qquad (5)$$

The function $F(N,a)$ in Eq.(4), which defines the net reproduction rate per individual from $a$-group, does not depend on other groups but only on total population size. It follows that the value of parameter $a$ is inherited inside an $a$-group. Although the last supposition is rather strong it is essentially less restrictive than usual supposition that *all individuals have identical value* of a reproduction rate (as in Eq.(2)).

### 2.2. Basic properties of inhomogeneous population models

The main assertions about the inhomogeneous population model (4), (5) are gathered in the following theorems 1-4.

Let $P_t(a) = l(t,a)/N(t)$ be the current *probability density function* (pdf) of the parameter $a$ and $E_t a$, $\sigma_t^2(a)$ be the mean value and variance over $P_t(a)$. Denote by $M_t(\lambda)$ the *moment generating function* (mgf) of $P_t(a)$:

$$M_t(\lambda) = \int_A \exp(\lambda a) P_t(a) da .$$

The mgf $M_0(\lambda)$ of the initial pdf $P_0(a)$ plays a fundamental role in the theory.

Let us introduce *auxiliary variables* $p(t)$, $q(t)$ by the system

$$dp/dt = g(N_0 q(t) M_0(p(t))), \tag{6}$$

$$dq/dt = qf(N_0 q(t) M_0(p(t))),$$

$$p(0)=0, \ q(0)=1.$$

**Theorem 1.** *Let Cauchy problem* (6) *has an unique solution* $\{p(t), q(t)\}$ *at* $t \in [0,T)$ *where* $0 \leq T < \infty$. *Then the functions*

$$l(t,a) = l_0(a) \, q(t) \exp((p(t)a), \tag{7}$$

$$N(t) = N_0 q(t) M_0(p(t)) \tag{8}$$

*satisfy system* (4), (5) *at* $t \in [0,T)$.

(ii) *Conversely, if* $l(t,a)$ *and* $N(t) = \int_A l(t,a)da$ *satisfy system* (4), (5) *at* $t \in [0,T)$, *then Cauchy problem* (6) *has a solution* $\{p(t), q(t)\}$ *at* $t \in [0,T)$ *and the given functions* $l(t,a)$, $N(t)$ *can be written in the form* $l(t,a) = l_0(a) q(t) \exp((p(t)a)$, $N(t) = N_0 q(t) M_0(p(t))$ *at* $t \in [0,T)$.

Theorem 1 shows that problem (4)-(5) is equivalent to Cauchy problem (6); it reduces problem (4)-(5) of population dynamics with distributed parameter to the classical Cauchy problem. Theorem 2 describes the dynamics of the main statistical characteristics of model (4)-(5).

**Theorem 2.** *Under conditions of Theorem* 1,

(i) *the current population size* $N(t)$ *satisfies the equation*

$$dN/dt = N[f(N) + E_t a \, g(N)] \tag{9}$$

(ii) *the current mean value of the parameter,* $E_t a$, *is determined by the formula*

$$E_t a = [dM_0(\lambda)/d\lambda \big|_{\lambda = p(t)}] / M_0(p(t));$$

*and satisfies the equation*

$$dE_t a /dt = g(N) \sigma_t^2(a), \tag{10}$$

(iii) *the current parameter variance,* $\sigma_t^2(a)$, *is determined by the formula*

$$\sigma_t^2(a) = d^2 M_0(\lambda)/d\lambda^2 \big|_{\lambda=p(t)}/M_0(p(t)) - ([dM_0(\lambda)/d\lambda\big|_{\lambda=p(t)}]/M_0(p(t))^2$$

*and is connected with the current mean by the relation*

$$\sigma_t^2(a) = d(E_t a)/dp. \tag{11}$$

Let us underline that the inhomogeneous model (4) governs not only dynamics of the total size $N(t)$ of the population, but also the evolution of the parameter distribution $P_t(a)$. This aspect of the inhomogeneous models is essentially new in comparison to "regular", homogeneous, dynamical models. Sometimes the evolution of the parameter distribution is of principal concern (it is the case in mathematical genetics). The following theorem describes dynamics of the distribution of the model parameter.

**Theorem 3.** *Under conditions of Theorem* 1,

(i) *The current parameter distribution $P_t(a)$ is determined by the formula*

$P_t(a) = P_0(a) \exp((p(t)a)/M_0(p(t))$;

(ii) *The pdf $P_t(a)$ solves the equations*

$dP_t(a)/dt = P_t(a) [g(N(t)))(a - E_t a)]$ .

The following important corollary of Theorem 1 explains the evolution of the composition of the inhomogeneous population.

**Corollary 1.** Under conditions of Theorem 1,

$l(t,a_1) / l(t,a_2) = l_0(a_1)/ l_0(a_2) \exp(p(t)(a_1-a_2))$.

Therefore, if $p(t)$ approaches infinity with $t \to \infty$ (as a rule, it takes place for positive functions $g(N)$), then the evolution of a heterogeneous population leads to the fast replacement of individuals with smaller values of the parameter *a* by those with greater values of *a*, even though the fraction of the latter in the initial distribution is arbitrarily small. A more general assertion, the Haldane principle for inhomogeneous population models, was given in [8].

Inhomogeneous population model defines not only the dynamics of the total population size, but also the evolution of the parameter distribution.

**Theorem 4.** *Under conditions of Theorem 1, let us assume that the initial distribution of the parameter a is*

(i) *normal with a mean $a_0$ and variance $\sigma_0^2$; then the parameter distribution will be also normal at any $t \in [0,T)$ with the mean*

$E_t a = a_0 + \sigma_0^2 p(t)$

*and with the same variance $\sigma_0^2$;*

(ii) *Poissonian with a mean $a_0$; then the parameter distribution will be also Poissonian at any $t \in [0,T)$ with the mean*

$E_t a = a_0 \exp(p(t))$;

(iii) *$\Gamma$-distribution with the coefficients k, s, $\eta$:*

$P_0(a) = s^k(a - \eta)^{k-1} \exp[-(a - \eta)s ]/\Gamma(k)$,

*where $s, k > 0, -\infty < \eta < \infty$ and $a \geq \eta$; $\Gamma(k)$ is the $\Gamma$-function.*

*Define $T^* = \inf(t \in [0,T): p(t) = s)$, if such t exists, otherwise $T^* = T$. Then the parameter a will be $\Gamma$-distributed at any time moment $t < T^*$ with coefficients k, s- p(t), and $\eta$ such that*

$$E_t a = \eta + k/(s - p(t)), \quad \sigma_t^2(a) = k/(s - p(t))^2.$$

An important conclusion can be drawn relying on the last formulae: even arbitrary, but non-zero variance of the initial $\Gamma$-distribution of the growth factor gives rise to the model «blowing up», i.e. the population size and also the mean and variance of the population distribution approach infinity at a certain instant $T$.

The list of practically implemented distributions can be extended. For our purposes, an *exponential distribution* (a special case of $\Gamma$-distribution with $k=1$) and a *bounded exponential distribution* defined on the finite interval is of particular interest:

$$P_0(a) = V \exp(-sa), \tag{12}$$

where $0 \le a \le c = const$, $s > 0$ is the distribution coefficient, and $V = s/(1-\exp(-sc))$ is the normalization constant.

The mgf of the probability distribution (12) is

$$M_0(\lambda) = s/(s-\lambda) [1 - \exp(c(\lambda-s))] / [1-\exp(-sc)]. \tag{13}$$

**Statement 1.** *Under conditions of Theorem 1, let us assume that the parameter a takes values in the interval $[0,c]$ and the initial parameter distribution is the bounded exponential distribution (12). Then at any instant $t \in [0,T)$ the parameter distribution is also of the form (12) in the same interval $[0,c]$ with coefficient $s-p(t)$.*

**Remark 1.** Equations (9), (10) with $f(N)=0$ read

$$dN/dt = N g(N) E_t a \tag{9'}$$

$$dE_t a /dt = g(N) \sigma_t^2(a) \tag{10'}$$

and show that, under positive $g(N)$, the mean of the parameter increases with time. If, additionally, the function $g(N)$ is non-decreasing, then the size of inhomogeneous population increases hyper-exponentially.

**Remark 2.** Equation (10') under $g(N)=1$ is well known in mathematical genetics as the Fisher's fundamental theorem of natural selection [3]. Its formulation and derivation, at least for asexual populations (see, e.g., [11], ch.4), show that this theorem is a simple but general statement about inhomogeneous populations, which is not specific to population genetics. Theorems 1-4 allow us to investigate the evolution of a trait distribution in more detail in many important cases if the initial distributions (see [2]).

2.4. Inhomogeneous Malthusian model

Let us consider the elementary but important model of population dynamics, the inhomogeneous Malthus model. Even this simplest model can show a surprising diversification of dynamic regimes of behavior subject to the initial distribution of the Malthusian parameter. The model is given by the equation

$$dl(t,a)/dt = al(t,a) \tag{14}$$

where $a$ is the Malthusian parameter. We suppose here that each individual has its own reproduction coefficient $a$, which is distributed over the population, and its distribution at the initial instant is $P_0(a)$ with mgf $M_0(\lambda)$.

The auxiliary variables $p(t)$, $q(t)$ for the inhomogeneous Malthus model are defined by equations (6), are $p(t)=t$, $q(t)=1$. Theorem 1 implies that, for all $t>0$ such that $M_0(t)$ exists, the following equalities are valid and define the model dynamics completely:

$$N(t) = N_0 M_0(t), \tag{15}$$

$$dN/dt = E_t a\, N,$$

$$P_t(a) = P_0(a)\exp(ta)/M_0(t).$$

In particular, if the parameter $a$ is $\Gamma$-distributed at the initial moment, then $M_0(t)= \exp(\eta t)/(1-t/s)^k$ (with $t<s$), and theorems 1,2 imply

**Statement 2.** *Consider the Malthus inhomogeneous model* (14) *and assume that the parameter a is $\Gamma$-distributed with the coefficients k, s, $\eta$ at the initial point in time. Then*

$$N(t) = N(0)\exp(\eta t)/(1-t/s)^k \text{ (for } t<s\text{)}, \tag{16}$$

$$E_t a = \eta + k/(s-t),\ \sigma_t^2(a) = k/(s-t)^2 \text{ for all } t<s.$$

In particular, if the initial distribution of the Malthusian parameter $a$ is exponential with $\eta=0$, then

$$N(t) = N(0)(1-t/s)^{-1} \text{ (for } t<s\text{)}. \tag{17}$$

Conversely, if the inhomogeneous Malthus model has solution (17) or (16), then the Malthusian parameter was accordingly exponentially or $\Gamma$-distributed at the initial moment.

Let us suppose now that the Malthusian parameter $a$ of model (14) takes values in a *bounded interval* [0,$c$] according to the bounded exponential distribution. Combining Statement 1 and Theorem 1, we arrive at

**Statement 3.** *Let the initial distribution of the Malthusian parameter a is the bounded exponential distribution (12). Define the time moment T=s. Then*

(i) *the current population size N(t) is defined by the formula*

$$N(t) = N(0) (1-t/s)^{-1}[1 - \exp(c(t-s))]/[1-\exp(-sc)] \qquad (18)$$

*and satisfies the equation* $dN/dt = E_t a \, N$, *where*

$E_t a = 1/(s-t) + c/[1-\exp(c(s-t))]$;

(ii) *at* $t \to T$, $N(t) \to N(s) = N(0) \, sc/[1-\exp(-sc)]$, $E_t a \to c/2$;

(iii) *after the moment T*

$N(t) = N(0) (t/s -1)^{-1}[\exp(c(t-s))-1]/[1-\exp(-sc)]$, *with* $t > s$,

$E_t a \to c$ *at* $t \to \infty$.

Therefore, the inhomogeneous Malthus model with bounded exponential distribution of the parameter is defined at any given point in time.

It was proved in [8] that the set of all functions $N(t)$ (15), which solve some inhomogeneous Malthusian model with positive initial parameters, coincides with the set of all absolutely monotonic functions, as opposed to the standard Malthus model, which has the unique solution $N(t) = N_0 \exp(at)$. This assertion demonstrates the broad scope of applications of non-uniform Malthus models. It also implies that modeling of inhomogeneous population dynamics on the basis of only the mean value of the reproduction rate without taking into account its distribution or at least variance is likely to be substantially incorrect. Indeed, even the dynamics of inhomogeneous Malthusian models with the same initial mean value of the Malthusian parameter can be very different depending on the initial distribution of the parameter (see, e.g., Theorem 4). Let us point out that all real populations are inhomogeneous.

### 2.5. Inhomogeneous logistic models

The theory of inhomogeneous models presented above can be extended to inhomogeneous models of populations that depend on many parameters $(a_1, ... a_n) = \boldsymbol{a}$. Description of the general theory is beyond the scope of this paper. Let us consider briefly only one important example, namely, the well-known Verhulst (or logistic) model with two parameters, which describes a self-limiting process of a population's growth:

$$dN/dt = aN(1-N/B), \qquad (19)$$

Here, the parameter *a* is equal to the initial reproduction rate of a very small population, the parameter *B* is the carrying capacity and the term $a(1-N/B)$ is the per capita net reproduction rate. The logistic model (19) can be written in the formally equivalent form

$$dN/dt = N(a_1 - a_2 N).$$

Here, the parameter $a_1$ is interpreted as birth rate and $a_2 N$ as death rate per capita, hence the death rate in this model increases when the population grows and approaches the stable size $N_{st} = a_1/a_2$. This is not the case when demographic processes are under consideration, in contrast with the model (19). In both versions, one of the parameters could be distributed and the other one fixed; the case of two distributed parameters is also of interest.

The theory of the inhomogeneous logistic model with the distributed parameter *a* under a fixed parameter *B* can be easily reduced to the theory of the inhomogeneous Malthus model. In brief, let us write equation (19) in the more general form

$$dN/dt = aNg(N) \tag{19'}$$

and consider the inhomogeneous model

$$dl(t,a)/dt = l(t,a)\, ag(N), \tag{20}$$

$$N(t) = \int_A l(t,a)\,da$$

with the parameter *a* having initial distribution $P(0,a)$ and corresponding mgf $M_0(\lambda)$.

Let $p(t)$ be a solution of Cauchy problem

$$dp/dt = g(N_0 M_0(p)),\ p(0)=0, \tag{21}$$

at $t \in [0,T)$ where $0 \leq T < \infty$. Then it follows from theorems 1-3, that $N(t)$ solves the equation

$$dN/dt = E_t a\, N\, g(N), \tag{22}$$

and is determined by the formula

$$N(t) = N(0) M_0(p(t)).$$

The current mean value $E_t a$ satisfies the equation

$$dE_t a/dt = g(N)\sigma^2(t)$$

and is determined by the formula

$$E_t a = d[\ln M_0(\lambda)/d\lambda \big|_{\lambda=p(t)}].$$

Let us note that the equation (22) in the parametrical form reads $dN/dp = E_t aN$ and hence coincides with the corresponding equation for the Malthus inhomogeneous model under changing *t* on *p* (and the same is valid for other relations). We may consider the variable *p* as "internal time" of a population governed by system (20), which coincides with the "plain" time under $g(N)\equiv 1$. The dynamics of the inhomogeneous logistic model (20) with respect to the "internal time" is the same as the dynamics of the inhomogeneous Malthusian model with respect to the "plain" time.

In what follows we consider a logistic-type inhomogeneous model of the form [5]

$$dl(t,a)/dt = al(t,a)\,(1-(N/B)^r),\ r=\text{const}>0.$$

Let us note that Eq.(21) for $g(N)= 1-(N/B)^r$ has a unique stable (and attractive) point $p^*$ if (and only if) $N_0<B$; this point is defined by the relation $M_0(p^*)= B/N_0$ because the mgf $M_0(\lambda)$ increases when $\lambda$ increases, $M_0(0)=1$ and $dp/dt|_{t=0}>0$. It means that the Cauchy problem (21) has a global solution for all $t\in [0,\infty)$ and $p(t)$ tends to $p^* <\infty$ at $t\to\infty$. Hence, the limit state of the inhomogeneous logistic model (20) coincides with the current state of the inhomogeneous Malthus model at the instant $p^*$. In particular, the limit stable population size and the limit distribution of the parameter $a$ are

$$N^* = N_0 M_0(p^*),$$
$$P^*(a)= P_0(a)\exp((p^*a)/M_0(p^*). \tag{23}$$

Let $M_0(\lambda)$ be a mgf for the bounded exponential distribution (13). Then Eq. (21) reads

$$dp/dt = 1-\{N_0s/(s - p(t))[1 - \exp(c(p(t)-s))]/[(1-\exp(-sc))B]\}^r,\ p(0)=0.$$

Basing on Statement 1 and Theorems 1, 2 we formulate

**Statement 4.** *Let us assume that the parameter a of logistic c model (20) assumes values in the interval* $[0,c]$ *and the initial distribution of the parameter is the bounded exponential distribution* (12). *Let $p(t)$ be a solution of Cauchy problem*

$$dp/dt = 1-\{N_0s/(s - p)[1 - \exp(c(p-s))]/[(1-\exp(-sc))B]\}^r,\ p(0)=0.$$

*Then*

(i) *the population size $N(t)$ is defined by the formula*

$$N(t) = N(0)\,s/(s - p(t))[1 - \exp(c(p(t)-s))]/(1-\exp(-sc)); \tag{24}$$

(ii) *$N(t)$ satisfies the equation*

$$dN/dt = E_t a\, N(1- (N/B)^r);$$

(iii) *the mean value and the variance of the parameter at every moment t are equal to*

$$E_t a = 1/(s- p(t)) +c/[1-\exp(c(s - p(t)))]. \tag{25}$$
$$\sigma_t^2(a)=[d(E_t a)/dp] = 1/(s-p(t))^2 -c^2\exp(c(s-p(t)))/[1-\exp(c(s - p(t)))]^2.$$

(iv) *the limit parameter distribution and the total stable population size are determined by formulas (23), (24) where $p^*$ is a root of the equation*

$$M_0(p^*)= s/(s- p^*) [1- \exp(c(p^*-s))] / [1-\exp(-sc)] =B/N_0. \tag{26}$$

Let us emphasize a notable property of the inhomogeneous logistic model with a distributed Malthusian parameter: it stays inhomogeneous at any instant and has non-generated limit distribution of the parameter at $t \to \infty$.

The logistic inhomogeneous model

$$dl(t,\boldsymbol{a})/dt= l(t,\boldsymbol{a})(a_1-a_2 N)= a_1\, l(t,\boldsymbol{a})\, (1-a_2/a_1 N), \tag{27}$$
$$N(t) = \int_A l(t, a_1, a_2) da_1 da_2.$$

with two distributed parameters was studied in [1]. Authors proved that if the domain of parameters is a rectangle, $A=(a,b) \times (c,d)$, then the population density will concentrate in the course of time at the point $(b,c)$, where the ratio $a_1/a_2$ reaches maximum.

If the domain of parameter is not a rectangle, the asymptotic behavior of the two-parametric logistic model may be more complex. The solution of the inhomogeneous model (27) and the evolution of the distribution of the vector-parameter $\boldsymbol{a}$ can be explored in more detail by the same methods as were described above. Briefly, let $P(0,\boldsymbol{a})=P(0;a_1,a_2)$ be a joint initial distribution of the parameters $a_1$, $a_2$ with values in the domain $A$; define the generalized moment generating function of this distribution with the formula

$$M(\lambda_1, \lambda_2)= [\int_A (\exp(\lambda_1 a_1 - \lambda_2 a_2) P(0; a_1, a_2).$$

Let us assume that the Cauchy problem

$$dq(t)/dt=N(0)M(t, q(t)),\ q(0)=0$$

has a global solution in $[0,T)$. Then for all $t \in [0,T)$

$$l(t,\boldsymbol{a}) = l(0,\boldsymbol{a}) \exp(ta_1 - q(t)a_2),$$
$$N(t) = N(0)\, M(t, q(t)),$$
$$P(t,\boldsymbol{a}) = P(0,\boldsymbol{a}) \exp(ta_1 - q(t)a_2) / M(t, q(t)).$$

It follows from here that, for two $\boldsymbol{a}$-groups with different values of the vector-parameter, $\boldsymbol{a}^1$ and $\boldsymbol{a}^2$,

$$l(t,\boldsymbol{a}^1)/ l(t,\boldsymbol{a}^2) = l(0,\boldsymbol{a}^1)/ l(0,\boldsymbol{a}^2)\, \exp(t(a^1_1- a^2_1) - q(t)(a^1_2- a^2_2)).$$

Taking into account that the function $q(t)$ is positive and increases, we see that, if the domain of values of the parameters $a_1$ and $a_2$ contains a "boundary" point $\boldsymbol{a}^*=(a^*_1, a^*_2)$ in which the value of the parameter $a_1$ is maximum and the value of the parameter $a_2$ is minimum $l(t,\boldsymbol{a})/ l(t,\boldsymbol{a}^*) \to 0$ for any point $\boldsymbol{a} \neq \boldsymbol{a}^*$. Thus, in the course of time, the population tends to a homogeneous state and will consist of $\boldsymbol{a}^*$-group only (as proved rigorously in [1]). In general case the asymptotic behavior of the model solution may be more complex. In particular, if there exist many points with the largest ratio $a_1/a_2$ or if a functional dependence between parameters $a_1$, $a_2$ exists (e.g., $a_2=Ca_1$, as in model (19)) then the population may stay inhomogeneous even in the limit stable state.

### 3. Global demography models

#### 3.1. The inhomogeneous Malthus model and the hyperbolic growth

The global demography model should take into account the non-homogeneous character of world population, which consists of many groups and subpopulations with different values of reproduction factor. Thus, we shall construct a model by relying on the non-homogeneous population model (4)-(5). Bearing in mind that the model solution is in the form of hyperbola (1), let us notice that according to Eq. (6) the auxiliary variable $p(t)=t$ if and only if $g(N)\equiv 1$. So the desired model should look like $dl(t,a)/dt=l(t,a)[f(N)+a]$ with the solution in the form $N(t)=q(t)M_0(t)$ (see Eq. (8)), and $q(t)\equiv 1$ if only $f(N)\equiv 0$. It means that the desired model is Malthus one with distributed Malthusian parameter. According Statement 2 and equality (16) we arrive to the following important conclusion.

*Hyperbolic formula of growth* (1) *is the solution* (16) *of the inhomogeneous Malthus model* (14) *with $\Gamma$-distributed Malthusian parameter for $s=T$, $\eta=0$ and corresponding values of k.*

In particular, the hyperbola $N(t)=C/(T-t)$ is the solution (17) of the inhomogeneous Malthus model (14), whose initial distribution is exponential with the mean and variance equal to 1/2025 and the initial population size is $N(0)\cong 10^8$.

**Remark 3**. Varfolomeyev and Gurevich [14] suggested a demographic model of the form $dN/dt=kJ(t)N(t)$ with the growth rate $kJ(t)$ varying in time. The model with both variants of $J(t)$ (equations (9) and (10) of the cited paper) could be considered in the frameworks of inhomogeneous Malthus model; it corresponds to inhomogeneous Malthus model (14) with the exponential and Poissonian distribution of the parameter $a$ accordingly, see Theorem 4.

**Corollary 2.** *In models* (16) *and* (17), *the moment T of «demographic explosion» is determined by the initial variance of the Malthusian parameter* $\sigma_0(a)$, *namely* $T=1/\sigma_0(a)$.

Suppose that the world population size grows according to model (16). Assume that it is possible to change the coefficients $\eta$ and $s=1/\sigma_0$ of the exponential distribution of the Malthusian parameter by governing the demographic policy. Then it is easy to see from Statement 2 that the moment of demographic explosion can be moved arbitrarily far without reduction (and even under increase) of an average reproduction rate by only *reducing its current variance.*

This conclusion apparently contradicts the common point of view on the problem of demographic explosion. It implies (if model (16) holds) that the main efforts should be directed not towards the restriction of the birth rate in the poorest regions, but towards smoothing life conditions, which consequently would decrease the *variance* of the reproduction coefficient relevant to the humankind as a whole.

*A numerical example.* Notice that in model (17) $E_{2000}\, a = \sigma_{2000}(a) = 0.04$. To shift the moment of demographic explosion to the year 2100 without reducing the average value of reproduction coefficient, it would suffice to change the numerical factors entering the Malthusian-parameter exponential distribution beginning from 2000 as follows: $\eta=0.03$, $\sigma_0=0.01$. Then, the new date of demographic explosion will fall on $T_1 = 2000 + 1/\sigma_0 = 2100$.

Furthermore, $N(t)=N(2000)\exp(\tau\eta)/(1-\tau\sigma_0)$, where $\tau = t-2000$. Then it follows from (16) that $N(2025) = N(2000) \exp(0.75)/0.75 = 2.823\, N(2000)$, $E_{2025}\, a = \eta+\sigma_0/(1-25\sigma_0) = 0.0433$, and $\sigma(2025) = 0.0133$.

The considered example shows that, although the "demographic explosion" could be moved arbitrarily far, it is inevitable within the framework of model (16); thus, this model, although fitting well the past growth of the humankind, is definitely wrong when applied to the future and requires replacement or modification.

3.2. How can the problem of «Demographic explosion» be eliminated from the models?

To make clear which model assumptions give rise to the unrealistic prognosis of «demographic explosion», let us consider the non-uniform Malthus model with the more realistic *bounded* exponential distribution (12) of the parameter. Then the «demographic explosion» disappears from the model. Indeed, according to Statement 3, the inhomogeneous Malthus model with bounded exponential distribution of the parameter having solution (18) has a meaning at any point in time, as distinguished from models (1), (2) and (16), (17). It is clear now that the «demographic explosion» is a corollary of the unrealistic supposition that the individual birth coefficient may take *unlimitedly large* values with positive probabilities. (This supposition is implicitly incorporated into formula (1) and is presented explicitly in models (2) and (16)). To fit the demographic data for the last several thousands years (with the accuracy about 10% of the relative mean-square deviation) one should take $c \approx 0.114$. Taking this value of $c$ and $s=2026$, with a very high accuracy one finds $[1- \exp(-sc)]=1$, so that Eq. (18) can be replaced by

$$N(t) = N(0) (1-t/s)^{-1}[1 - \exp(c(t-s))] \qquad (28)$$

or for arbitrary initial time moment $t_0$

$$N(t) = N(t_0) (s-t_0)/(s-t)[(1 - \exp(c(t-s))/ (1 - \exp(c(t_0-s)))] \ .$$

Fig. 1 shows the evolution of the distribution of the Malthusian parameter $a$: the initial truncated exponential distribution, with time, concentrates at the point $c$, the maximal possible value of the parameter (according Corollary 1). This process explains the most important feature of model (18), which describes the change of the initial *hyperbolic* growth of $N(t)$ to the final stage of growth, which approaches *exponential* growth.

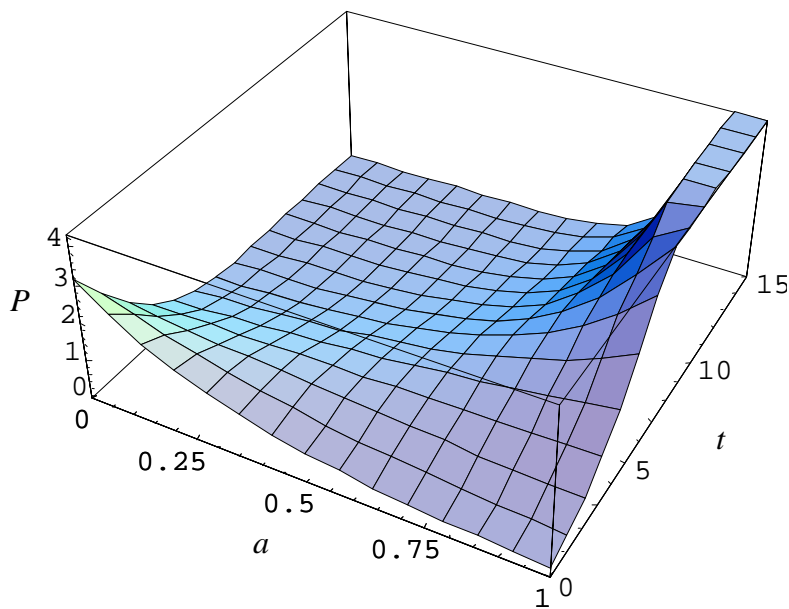

Fig.1. Evolution of the initial truncated exponential distribution of Malthusian parameter $a$ in time; $P$ – the current pdf of the parameter $a$ at time moment $t$.

On the other hand, this result allows us to suppose that, perhaps, there is no dramatic difference between the growth of human population and that of other biological populations in a free regime. Indeed, let us suppose that a population grows according to the inhomogeneous Malthus model with any *bounded* initial distribution of the Malthusian parameter. Then, according to Corollary 1, the current parameter distribution concentrates in the course of time at the maximal possible value of the parameter and the model shows exponential growth. The initial growth is accomplished through increasing the mean reproduction rate and hence is "faster" then the exponential one, see equation (9').

The transition from the initial "hyper-exponential" to the "almost exponential" phase of development may take comparatively little time (under short reproduction age and/or corresponding properties of the initial distribution, such as large variance of the initial exponential distribution). Hence, this transition may be unrecognized for some populations. As for the humankind, its development during the entire historic period until the last decades of the 20[th] century followed the hyper-exponential regime.

### 3.3. Logistic inhomogeneous model

Although formula (28) is essentially more appropriate than initial models (1) or (16), it predicts unlimited growth of $N(t)$ when $t \to \infty$. So, instead of Malthusian inhomogeneous model let us consider a logistic-type inhomogeneous model [5]

$$dl(t,a)/dt = a\, l(t,a)\, [1 - (N/B)^r], \quad r=\text{const}>0 \tag{29}$$

that takes into account the upper boundary of the possible population size with the help of the parameter $B$, such that $N=B$ is the stable state of (29).

Suppose that the upper boundary for the total population size is fixed and the parameter $a$ is distributed; suppose also that it has the same bounded exponential distribution at the initial instant as the Malthusian model considered in s.3.2. Then the dynamics of the total population size and the evolution of the parameter distribution are completely described in Statement 4.

The solutions of three considered models are shown in Fig.2.

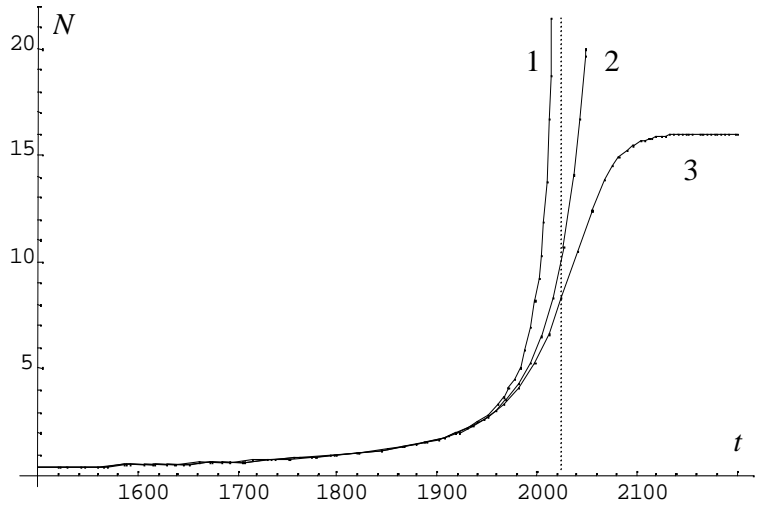

Fig.2. World population size, $N$ (in bln), dependently on time $t$: 1-models (1) and (17) with $s=2025$; 2 - model (28) with $c=0.114$; 3-model (29), (24) with $B=16$, $r=1.5$.

It is also interesting to compare the dynamic behavior of the mean $E_t a$ of the Malthusian parameter $a$ for different models as shown in Fig. 3.

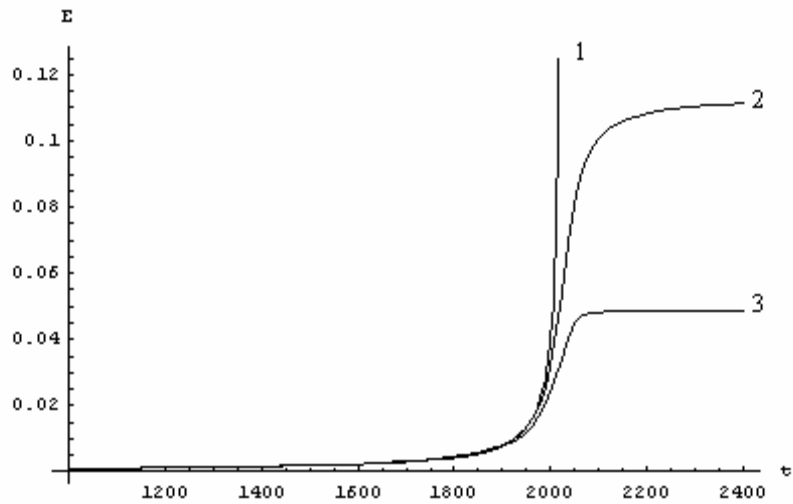

Fig.3. Mean value $E$ of the parameter $a$ dependently on time $t$: 1 - models (1) and (17) with $s=2025$; 2 - model (28) with $c=0.114$; 3 - model (29), (24) with $B=16$, $r=1.5$.

Finally, let us compare the behaviors of the variance $\sigma_t^2(a)$ of the parameter $a$ for different models (formula (11) gives the simplest way to compute $\sigma_t^2(a)$).

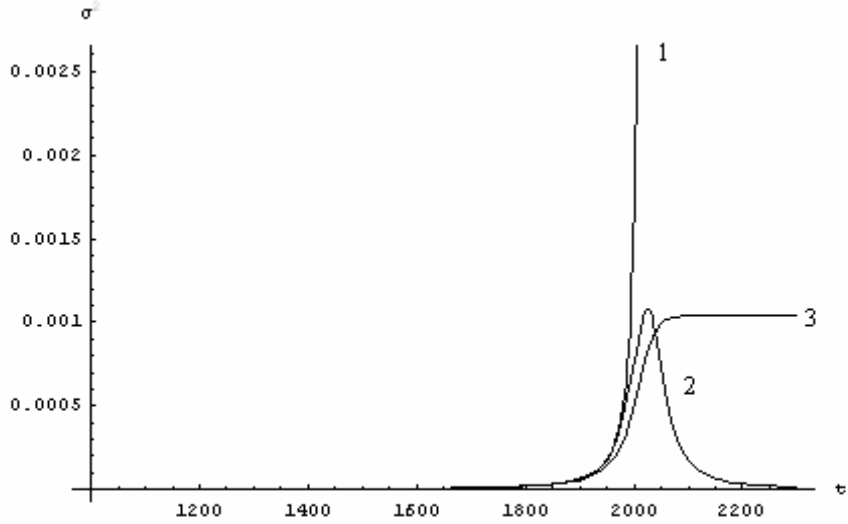

**Fig.4.** Variances $\sigma_t^2$ of the parameter *a* dependently on time *t*: 1 - models (1) and (17) with *s* =2025; 2 - model (28) with *c* = 0.114; 3 - model (29), (24) with *B*=16, *r*=1.5.

Figures 3 and 4 show that the evolution of the distribution of the parameter *a* is dramatically different for different considered models. For the Malthusian model with exponentially distributed parameter (model (14) with solution (1)), both the mean and the variance of the Malthusian parameter diverge as well the total population size at the moment T≈2025.

For the Malthusian model with *truncated* exponential distribution of the parameter *a* in the interval [0,*c*] (model (14) with solution (18) or (28)), the mean $E_t a$ is finite at the moment *T*, $E_T a = c/2$, and $E_t a$ tends to the maximal possible value, *c*, at $t \to \infty$ (see Statement 3). It means that, after the moment *T*, the model shows growth which approaches *exponential*. The variance $\sigma_t^2(a)$ vanishes at $t \to \infty$, hence the population becomes homogeneous. The transition period is clearly marked by a sharp ("impulse"-like) increase and then decrease of the variance of the Malthusian parameter.

For the logistic model with the truncated exponential distribution of the growth parameter *a* (model (29) with solution (24)), the mean value $E_t a$ increases monotonously to a limit value that depends on the model parameters *c, s, B* (see Statement 4). More precisely, the limit value $E^*$ of $E_t a$ is defined by formula

$E^* = 1/(s - p^*) + c/[1 - \exp(c(s - p^*))]$

where $p^*$ solves equation (26). Due to Statement 4, the variance $\sigma_t^2(a)$ at $t \to \infty$ tends to

$$\sigma^2{*}=1/(s-p^*)^2-c^2\exp(c(s-p^*))/[1-\exp(c(s-p^*))]^2>0.$$

Thus, the logistic model stays inhomogeneous at all times, even when the total population size is stabilized. Note that an individual reproduction coefficient, which is equal to $a[1-(N/B)^r]$, as well as the net growth factor $E_t a[1-(N/B)^r]$, tend to zero as the total population size $N$ approach the limit value $B$.

To demonstrate the potential of the developed approach, let us compare the solutions of the inhomogeneous logistic model (20)-(25) with some global demographic data and forecasts, namely: the IIASA data (Fig. )5, the data and forecast of the U.S. Bureau of the Census, International Data Base (Fig. 6), and the data and forecast of UNO [2] (Fig. 7).

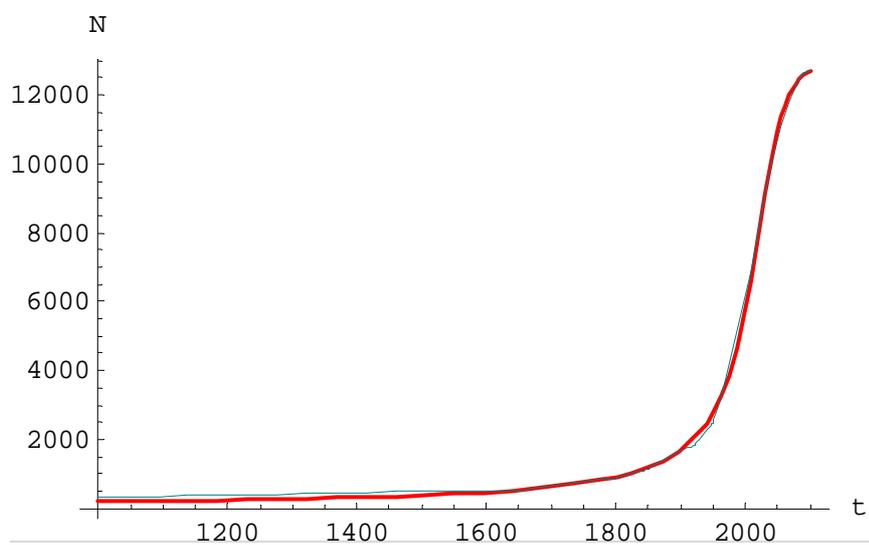

Fig. 5. The IIASA data and forecast of the world population size, $N$ (thin); the solution (thick) of the logistic model with $c=0.114$, $s=1976.6$, $r=1.08$, $B=12900$, and $N(0)=0.094$.

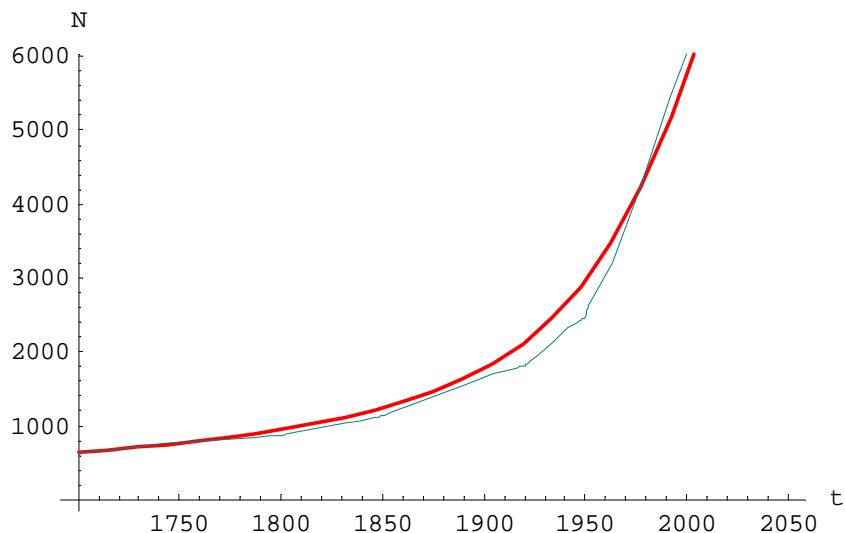

Fig. 6. The U.S. Bureau of the Census data and forecast of the world population size, *N* (thin); the solution (thick) of the logistic model with *c*=0.114, *s*=1976.6, *r* =1.26, *B*=9200, and *N*(0)=0.094.

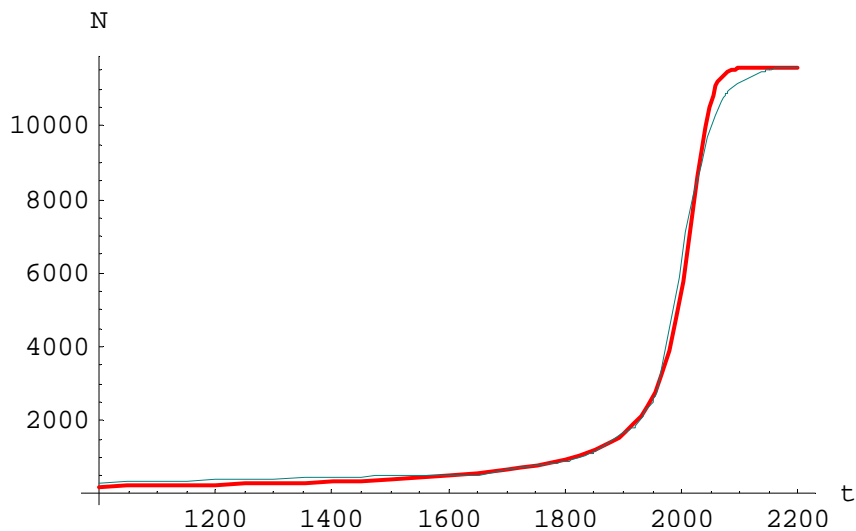

Fig. 7. The UNO data and forecast of the world population size, *N* (thin); the solution (thick) of the logistic model with *c*=0.114, *s*=2026, *r*=1.8, *B*=11600, and *N*(0)=0.104.

These examples are given mainly to demonstrate the capabilities of the developed approach and constructed models to fit real demographic data and existing forecasts. Conceivably, the parameter values and the accuracy of solutions may be improved. Nevertheless, it is of interest that the "critical" year corresponding to the inflection point of *N*(*t*) falls on *T*=*s*=1976 for the logistic non-uniform model and forecasts of IIASA and the USA BC. This conclusion agrees with real data but differs essentially from the estimates made by the hyperbolic model (*T*=2025) as well as the non-uniform Malthusian model and the earlier forecasts of UNO (*T* =2023).

**Discussion and perspectives**

The main assumption of our approach to modeling of a population's dynamics is that every individual has its own, specific value of the reproduction coefficient. We show that, in an inhomogeneous population with a distributed reproduction coefficient, the entire initial distribution of the coefficient should be used to investigate the real population dynamics. Knowledge of the initial mean value and variance or even any finite number of moments of the initial distribution may be insufficient to predict the overall population growth.

We show that an individual-based inhomogeneous population model at a given initial distribution is equivalent to a Cauchy problem (defined for each specific model) and can be investigated completely. We examined the population growth and the evolution of the distribution of the reproduction coefficient, in particular, the dynamics of its mean value and variance. Any changes of the mean value, variance and other characteristics of the parameter distribution with time are caused only by variation of the population structure.

We derived three conceptual models of the population growth using recent developments in the theory of inhomogeneous population dynamics. The first model, an inhomogeneous Malthus model with exponentially distributed Malthusian parameter $a$ (the reproduction coefficient per individual), suggested a new explanation of the well-known hyperbolic law of the human population growth over long time intervals. The *demographic explosion,* a paradoxical prognosis of the hyperbolic model, is a consequence of an implicit, unrealistic assumption that the individual reproduction rate ($a$) may have arbitrarily *large* values with positive probabilities. A moment $T$ of the demographic explosion is defined by the initial variance of the parameter $a$. Additionally, the mean value and variance of the parameter $a$ also diverge at the moment $T$.

The second model, an inhomogeneous Malthus model with *truncated* exponentially distributed Malthusian parameter $a$ (concentrated in any interval $[0,c]$), eliminates the *demographic explosion.* The model solution shows a remarkable phenomenon, the transformation of the protracted hyperbolical growth to an exponential growth. This transformation is caused by the dynamics of the distribution of the reproduction coefficient due only to inner variation of the population structure. The transition period is rather short and is characterized by a sharp (bell-shaped) growth of the variance of the parameter $a$. After this period, the parameter variance tends to 0, whereas its mean value tends to $c$, the maximal possible value.

Both models describe the following process of variation of the population structure: individuals having the larger reproduction coefficients replace, with time (exponentially fast), those with smaller values of this parameter.

Finally, we consider the logistic modification of the second model. The solution to the logistic model with a *truncated* exponentially distributed parameter $a$, the birth rate per individual, is bounded and shows the transition from protracted hyperbolical growth to saturation. The transition period is also rather short and is characterized by convergence of the mean and variance of the parameter *a* to the corresponding *non-zero* stable values.

The latter model allows us to fit accurately the world population data for the last 2000 years as well as some of the existing forecasts (UNO and IIASA). The model predicts that the population will be inhomogeneous indefinitely and that *every* subpopulation present at the initial instant will be present in the limit state. In contrast, the logistic model, in which both independent parameters are distributed, predicts that a single subpopulation will be present in the limit state.

The logistic model (27) with both distributed parameters $a_1$, $a_2$ could also be applied to demographic modeling but it seems that the parameters of the model cannot be independent. Indeed, it follows from the results of [1] that, in the case of independence, the whole population tends to become homogeneous and all subpopulations (except for one) disappear exponentially. This prognosis looks unrealistic. The opposite case, $a_1 \sim a_2$, was studied above (model (29)). This model predicts that *every* subpopulation will be presented in the limit stable state and this prognosis might not be entirely realistic. Investigation of the inhomogeneous Allee-type model with the "lower critical number of a population" is a promising next step. An interesting problem of revealing stochastic dependence between parameters of the inhomogeneous logistic and Allee-type models remains open.

The results of analysis of inhomogeneous population models presented here suggest that the development of human population follows the same laws as other biological populations, a notion that, perhaps, contradicts the common point of view (e.g., [10], ch.-s 7, 21]). The initial growth of any inhomogeneous population includes the "hyper-exponential" phase caused by the increase of the mean reproduction rate.

We could hypothesize that the Malthusian growth of a population with a constant reproduction rate is not an initial "free" regime but rather a transition from the initial hyper-exponential growth, when the net reproduction rate increases, to the final stage, when the net reproduction rate decreases and tends to zero. The hyper-exponential growth phase is inherent to the development of any inhomogeneous population.

The transition from the initial "hyper-exponential" to "almost exponential" development might take comparatively little time and hence could go unrecognized for some populations. As for the humankind, its development during the entire historic period followed the hyper-exponential phase characterized by hyperbolic growth. This phase came to an end only in the last years of the 20$^{th}$ century; now (2004) we are at the transition period from the "almost exponential" to the saturation regime. The difference between the humankind (considered as a heterogeneous population) and other biological populations is that the former stayed the initial phase of growth for an unusually long time, perhaps due to an extremely successful struggle against the equalizing pressure of the natural selection.

**Acknowledgements**

I thank Prof.-s O. Diekmann, J. Cushing, C. Castillo-Chavez, L. Gross, and T. Hallam and participants of seminars in Tennessee University, Cornell University and Utrecht University for useful discussions; I thank also F. Kondrashov, Dr. Y. Kim and Dr. E. Koonin for valuable comments and help in preparation of the manuscript.